\documentclass[useAMS,usenatbib]{mn2e}

\usepackage{graphicx}
\usepackage{longtable}
\usepackage{subfigure}
\usepackage{url}

\title[The Simultaneous Medicina-{\it Planck} Experiment]{The Simultaneous Medicina-{\it Planck} Experiment (SiMPlE): data acquisition, reduction and first results}
\author[P.~Procopio et al.]{
\parbox[t]{\textwidth}
{P.~Procopio$^{1}$\thanks{E-mail: procopio@iasfbo.inaf.it},
M.~Massardi$^{2}$, S.~Righini$^{3}$, A.~Zanichelli$^{3}$,
S.~Ricciardi$^{1}$, P.~Libardi$^{3}$,
C.~Burigana$^{1,6}$, F.~Cuttaia$^{1}$, 
K.-H.~Mack$^{3}$,   L.~Terenzi$^{1}$, F.~Villa$^{1}$,
L.~Bonavera$^{4,8}$, G.~Morgante$^{1}$,
C.~Trigilio$^{5}$, T.~Trombetti$^{1,7}$, G.~Umana$^{5}$}
\vspace*{8pt} \\
$^{1}$ INAF-IASF Bologna, Via Gobetti 101, I-40129 Bologna, Italy\\
$^{2}$ INAF - Osservatorio Astronomico di Padova, Vicolo dell'Osservatorio 5, I-35122 Padova, Italy\\
$^{3}$ INAF-IRA Bologna, Via Gobetti 101, I-40129, Bologna, Italy\\
$^{4}$ SISSA/ISAS, Via Bonomea 265, IÐ34136 Trieste, Italy\\
$^{5}$ INAF - Osservatorio Astrofisico di Catania, Via S. Sofia 78, I-95123 Catania, Italy\\
$^{6}$Dipartimento di Fisica, Universit\`a degli Studi di Ferrara, Via G. Saragat 1, I-44100 Ferrara, Italy\\
$^{7}$ Dipartimento di Fisica, Universit\`a La Sapienza, P.le A. Moro 2, I-00185 Roma, Italy\\
$^{8}$ Australia Telescope National Facility, CSIRO Astronomy and Space Science, PO Box 76, Epping, NSW 1710, Australia}
\begin{document}

\date{}

\pagerange{\pageref{firstpage}--\pageref{lastpage}} \pubyear{}

\maketitle

\label{firstpage}

\begin{abstract}

The Simultaneous Medicina-{\it Planck} Experiment (SiMPlE) is aimed at observing a selected sample of 263 extragalactic and Galactic sources with the Medicina 32-m single
dish radio telescope in the same epoch as the Planck satellite observations.

The data acquired with a frequency coverage down to 5 GHz, also combined with Planck at frequencies above 30 GHz, will constitute a useful reference catalogue of bright sources over the
whole Northern hemisphere. Furthermore, source observations performed in different epochs and comparison with other catalogues allow the investigation of source variabilities on different timescales.

In this work, we describe the sample selection, the on-going data acquisition campaign, the data reduction procedures, the developed tools, and the comparison with other data-sets.

We present the data at 5 and 8.3 GHz for the SiMPlE Northern Sample consisting of 79 sources with $\delta \geq 45^\circ$ selected in our catalogue and observed during the first
6 months of the project.
A first analysis of their spectral behaviour and long-term variability is also presented.

\end{abstract}


\section{Introduction}

Microwave band observations with satellites mainly aimed at the study of the cosmic microwave background (CMB) also offer a unique opportunity to investigate the spectral properties of radio sources in a poorly explored frequency range,
partially inaccessible from ground.

The study of the Spectral Energy Distribution (SED) and variability of sources from radio to far-infrared (FIR) wavelengths is crucial to probe the physics of the innermost regions of Active Galactic Nuclei (AGN) and radio galaxies and of some classes of variable Galactic radio sources (i.e. microquasars and binary active systems). These objects show remarkable outbursts and studying their SED simultaneously from radio to FIR in several epochs helps constraining the outburst mechanisms and the processes that activate their radio emission.  

The project described in this paper aims at accumulating multifrequency radio observations of the northern sky at 5, 8.3 and 22
GHz almost simultaneous with those taken by the ESA {\it Planck} satellite (Tauber et al. 2010a) from 30 to 857 GHz. This will allow the construction of the SED of a significant sample of sources on a wide frequency range through almost coeval data. The simultaneity of the observations is crucial to evaluate the contribution from variability and spectral effects in affecting the physical interpretation of source observational properties.

Catalogues based on the Wilkinson Microwave Anisotropy Probe (WMAP) satellite maps have been produced using several source extraction techniques (L\'opez-Caniego et al. 2006; Chen \& Wright 2009; Bennett et al. 2003; Hinshaw et al. 2007; Gold et al. 2011). Their completeness limit (i.e. the flux density level above which all the sources in the observed area are listed in the catalogue) is typically $\sim 1\,$Jy at 23 GHz.
Massardi et al. (2009) have combined blind and non-blind approaches (see L\'opez-Caniego et al.\ 2007, Gonz\'alez-Nuevo et al.\ 2008) exploiting the 
Mexican Hat Wavelets 2
(MHW2) filter (Gonz\'alez-Nuevo et al.\ 2006) to extract an all-sky catalogue of 516 sources with $|b|>5^\circ$. It is almost 91 per cent complete above 1 Jy and constitutes the New Extragalactic WMAP Point Sources (NEWPS) catalogue. Almost all (484) sources in the sample were previously catalogued as extragalactic (457) or Galactic (27) objects. The remaining 32 candidate sources do not have counterparts in lower-frequency all-sky surveys with comparable flux densities and may therefore be just high peaks in the distribution of other components present in the maps. If they are all spurious, the reliability of the sample (i.e. the probability that a source listed in the sample is a genuine radio source) is 93.8 per cent.

To date, the Australia Telescope 20-GHz Survey (AT20G, Murphy et al. 2010, Massardi et al. 2010) provides the deepest complete ground-based sample of the high-frequency Southern sky.
It is a blind survey performed with the Australia Telescope Compact Array (ATCA) in the years $2004 - 2008$. 
Its final catalogue consists of 5890 sources above a flux limit of 40 mJy and is an order of magnitude larger than any previous catalogue of high-frequency radio sources.
No analogous sample, deeper than the NEWPS catalogue has been observed at frequencies above 8.4 GHz in the Northern hemisphere.

The ESA {\it Planck} satellite (Tauber et al. 2010a)
is providing a blind survey of the entire sky in nine frequency bands
(30, 44, 70, 100, 143, 217, 353, 545, 857 GHz), with Full Width Half Maximum (FWHM) resolution ranging from 33 to 5 arcmin (Mennella et al. 2011, {\it Planck} HFI Core Team et al. 2011) and at several epochs.
Leach et al. (2008) estimated the possibilities of the MHW2-filter-based detection techniques applied to the {\it Planck} satellite maps, considering two all-sky surveys. They found that the expected detection limits range from $\simeq 0.4\,$Jy at 30 GHz to $\simeq 0.22\,$Jy at 100 GHz, which implies that all the NEWPS sources should be detectable also in the {\it Planck} maps up to $\sim100$ GHz.

The {\it Planck} Early Release Compact Source Catalogue (ERCSC) has been recently released (The {\it Planck} Collaboration 2011a) on the basis of more than one full coverage of the entire sky. The catalogue refers, in fact, to a stage when a first full-sky map plus a second run on 60 per cent of the sky were available.
A Monte-Carlo algorithm was implemented to select reliable sources
among all the extracted candidates such that the cumulative reliability of the catalogue is $\ge 90$ per cent. There is no requirement on completeness for the ERCSC.
The 10$\sigma$ photometric flux density limit of the catalogue at $|b| > 30^\circ$ is 0.49, 1.0, 0.67, 0.5, 0.33, 0.28, 0.25, 0.47 and 0.82 Jy at each of the nine frequencies between 30 and 857 GHz.

The {\it Planck} ERCSC  
``provides a robust list of stars with dust shells, stellar cores, radio galaxies, blazars, infrared luminous galaxies, Galactic interstellar medium features,
915 cold molecular cloud core candidates, 189 Sunyaev-Zel'dovich cluster candidates as well as unclassified sources'' (The {\it Planck} Collaboration 2011a).
Many of them are object of dedicated papers
about Galactic science, extragalactic sources, and Sunyaev-Zel'dovich effects and cluster properties\footnote{ The {\it Planck} Early Results papers are publicly available at the following web address \url{http://www.sciops.esa.int/index.php?project=PLANCK&page=Planck_Published_Papers}. }. 
 
The source list, with more than 15000 unique sources, is ripe for follow-up characterisation with Herschel and several ground-based observing facilities.

The first version of the {\it Planck} Legacy Catalog, 
together with other {\it Planck} products and a first set of cosmological papers, 
will be released in the first months of 2013, 
i.e. at the end of the proprietary period of the data from the first two surveys (namely, 15 months of observation). 

{\it Planck} observations help identifying the properties of the SED in the high frequency radio and FIR bands (The {\it Planck} collaboration 2011b, c), and benefit from the ground-based observations at longer wavelengths for the reconstruction of the properties of different emission components to study the spectral behaviour of different radio source populations .

An analysis of the variability on the five-year WMAP point sources shows that a high fraction of the sources are variable at more than 99 per cent  confidence,
and these are in general the brighter ones (Wright et al. 2009).
In general, most AGNs are variable at these radio frequencies. Their long-term variability has been well studied over the years at similar or higher frequencies, for example, by the 
University of Michigan Radio Astronomy Observatory (UMRAO) and 
Mets\"ahovi Radio Observatory research teams for sources in the northern sky and equatorial regions (Hovatta et al. 2007, Hughes et al. 1992).
Also, the comparison between the WMAP catalogue and the {\it Planck} ERCSC at the corresponding bands, that can be performed in a general, statistical sense because of source variability, 
reveals no systematic difference between the WMAP and ERCSC flux densities, while the significant scatter confirms
that variability is an issue (The {\it Planck} collaboration 2011a). Variability enhances the scattering of flux density comparisons.
It also results in a bias in favour of sources in a bright phase at the selection epoch. This will make some of the variable (and mostly bright flat spectrum) ones appear with 
a Gigahertz Peaked-Spectrum (GPS) like distorted spectrum characterising flaring phases.

Sadler et al. (2006) estimated a median debiased variability of 6.9 per cent at 20 GHz and on timescales of a year for a 100-mJy flux density limited sample of extragalactic radio sources, with only a few sources more variable than 30 per cent. Most recently, Massardi et al. (2011) estimated at 18~GHz a median debiased variability on 9 months of 9 per cent for a sample of AT20G sources with $S_{\rm 20 GHz}>$500 mJy.

The flux density levels achieved during outbursts by the above mentioned classes of Galactic radio sources are high enough to be observed with good signal-to-noise ratio (SNR) by {\it Planck}.

Driven by all the above reasons, several projects are carrying out observations simultaneous with the {\it Planck} satellite at various frequency bands. They involve observational facilities such the Australia Telescope Compact Array (ATCA), 
the Effelsberg Radiotelescope (Germany), the Institut de Radioastronomie Millim\'etrique 
(IRAM) 30 m Telescope (Spain) the Mets\"ahovi Radio Observatory (Finland), 
the Very Large Array / Expanded Very Large Array (VLA/EVLA) (USA),
operating in the frequency range between 4.5 and 40 GHz, 2.6 and 43 GHz, and at 86.2, 142.3 GHz, at 37 GHz, and at 5, 8, 22, 43 GHz, respectively (see The {\it Planck} Collaboration 2011b, c).

The  {\it Planck} ATCA Coeval Observations (PACO, Massardi et al. 2011) project is the largest one and has followed up a 
sample of 482 AT20G extragalactic sources in the frequency range between 4.5 and 40 GHz in the period between July 2009 and August 2010. 
Several sources were observed more than once to study variability.

We present here the Simultaneous Medicina {\it Planck} Experiment (SiMPlE), designed to complement the PACO project in the Northern hemisphere
using the Medicina 32-m single dish to observe a sample of 263 sources at 5, 8.3 and 22 GHz almost simultaneously to the {\it Planck} observations.

The SiMPlE data alone at low frequencies and combined with {\it Planck} for frequencies above 30 GHz, constitute a useful reference catalogue
of bright sources over the whole Northern hemisphere.
The sample selection criteria are described in Section \ref{sec:sample}, the observing strategy is given in Section \ref{sec:obsstrat}, and data reduction techniques are
discussed in Section \ref{sec:datared}.
Data and spectral analysis for the 79 NEWPS sources with $\delta \geq 45^\circ$ at 5 and 8.3 GHz are presented and discussed in Section \ref{sec:catalogue}.
Finally, the description of the present status of the project, our main findings,  and some future perspectives are summarised in Section \ref{sec:conclusions}.

\begin{figure}
 \centering
 \includegraphics[width=1.\hsize]{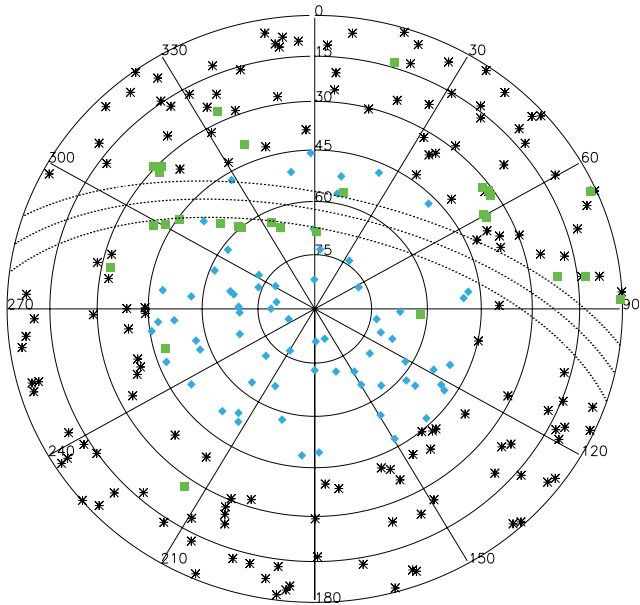}
 \caption{The SiMPlE sample: the Northern Sample (diamonds) covers the area with $\delta \geq 45^{\circ}$; asterisks show the Southern Sample; squares indicate the central positions of the mapped sky patches. The dotted lines represent a 10-degrees masked band centred on the Galactic plane (middle dotted line).
 }
\label{fig:map_srcs}
\end{figure}

\section{Sample selection}\label{sec:sample}

The selection of our sample of sources largely exploits the NEWPS sample (Massardi et al. 2009).
At the time of planning out observations it constituted a unique opportunity to select high-frequency
samples of bright sources in the Northern hemisphere, including 252 sources with declination $\delta>0^\circ$.

Cross-matches of the northern NEWPS sources with low-frequency wide-area surveys like the GB6 (Gregory et al.\ 1996) catalogue at 4.85 GHz or the NVSS (Condon et al.\ 1998) at 1.4 GHz have identified the candidate counterparts for 234 sources, 11 of which show multiple associations. Identifications based on the NASA Extragalactic Database (NED) classified 225 of them as extragalactic objects and the remaining 9 as Galactic. 18 sources do not have counterparts in lower-frequency all-sky surveys with comparable flux densities and may be spurious sources.

Because of the large WMAP beam and, as a consequence, of the positional uncertainty of the NEWPS detections, in planning follow-up observations we have used the coordinates of the low-frequency counterpart, where available, as target positions. In case of spurious sources or multiple associations we have chosen the WMAP positions as centre of a region
 that we have mapped at 5 GHz with the Medicina radio telescope in order to identify, in the case it was a compact radio source, the low-frequency counterpart of the WMAP detection (see Sect. \ref{sec:maps} for further details). The SiMPlE observations will therefore help to assess precisely the reliability of the detection techniques that have been applied to the WMAP maps.

Comparisons of the NEWPS sample detection with the AT20G Bright Source Sample (Massardi et al. 2008) data indicate that the NEWPS sample is almost 91 per cent complete down to 1 Jy. The completeness of the sample is determined by the flux density scales at the epoch of the WMAP observations, and variability can alter it. However, the comparison with the coeval 30-GHz {\it Planck} observations,
that are achieving deeper flux densities,  will allow us
to quantify the level of completeness of our sample at least at 22 GHz, once we will properly account for the difference in frequency.
The 22-GHz data and the comparisons with the {\it Planck} results will be presented in future papers from our group.

In addition to the Galactic sources in the NEWPS catalogue at $\delta>0^\circ$, the SiMPlE project includes the monitoring of a sample of 10 sources representative of various classes of highly variable objects, in case they may show a strong outburst during the {\it Planck} mission.
Massive X-ray binary systems, in fact, alternate quiescent periods and strong outbursts: about 10 per cent of the more than 200 known binary systems are radio-loud (Mirabel \& Rodrigues 1999). Some of them (e.g. Cyg X-3) are expected to reach flux densities up to 20 Jy (Trushkin et al., 2008; Szostek et al., 2008) at 8.4~GHz.
Luminous Blue Variable stars (e.g. Eta Carinae) show sudden outbursts during which the flux density can increase to several Jy at centimetre wavelengths and tens of Jy in the millimetre band. Finally, active binary stars (e.g. RS CVn and Algol) have periods of quiescence, with flux densities of a few tens of mJy, and active periods characterised by flares, lasting several weeks, every 2-3 months, reaching from several hundred mJy up to some Jy at cm wavelengths (Umana et al. 1995). Their radio spectra show the maximum of emission at frequencies higher than 10~GHz and, during the impulsive phases, this moves up to about 100 GHz.

Pre-emptive observations of these objects began in 2009 with the Noto telescope at 43 GHz and continued in Medicina, starting from June 2010 in the framework of the SiMPlE experiment at lower frequencies. These observations are on-going and will cover at least 1 years of {\it Planck} data acquisition. The results and the comparison with {\it Planck} data as well as 22-GHz data will be shown in future papers.

In summary, the SiMPlE sample contains 263 targets of which 253 are extracted from the NEWPS sample. 29 of these have been observed by mapping the surrounding region, while cross-On-The-Fly (OTF) scans have been used to observe the others. Fig. \ref{fig:map_srcs} shows the distribution on the Northern hemisphere of the observed sample. 79 objects have declination $\geq 45^\circ$ and constitute the so-called `Northern sample' for which in Section \ref{sec:catalogue} we present the results of the observations in the epochs between June and November 2010. The remaining epochs for the full sample will be discussed in future papers.

\section{Observing strategy with the Medicina radio telescope}\label{sec:obsstrat}

\subsection{Simultaneity with {\it Planck}}
\label{sec:simultaneity}

The {\it Planck} satellite scans the sky in circles passing close to the ecliptic poles and covers the whole sky in about 7 months
(Dupac \& Tauber 2005; see also section 1 of Mandolesi et al. 2010 and references therein). All the beams (Tauber et al 2010b) of the entire {\it Planck} field of view cross the same sky position in a few days at low and intermediate ecliptic latitudes, in weeks or few months at high ecliptic latitudes, and then pass again on it at the subsequent survey, so allowing to recover source flux density variations on different timescales (Burigana 2000, Terenzi et al. 2002, 2004). The extraction of the source flux density at {\it Planck} frequency channels is based on the analysis of frequency maps produced by optimally weighting the Time Ordered Data (TOD) in pixel space. Therefore, we can not achieve information from them on the very short term variability, a study that necessarily requires to exploit the information contained in time-ordered data, less sensitive than the channel maps.
We then consider observations to be `simultaneous' with the satellite if performed within 10 days from the satellite observations at any of its frequencies.
In practice, this has been also a reasonable compromise with the scheduled observational days available to the project.
This typical time sampling does not prevent the extraction of relevant variability information at least for the large majority of extragalactic sources, whose flux densities change appreciably on significantly longer timescales. The possibility to carry out observations, possibly simultaneously with {\it Planck}, of bright sources (in particular of objects expected to show faster variability)
with a finer timescale at the Medicina radio telescope is under investigation for the next SiMPlE campaigns.
In general, the {\it Planck} On-Flight Forecaster (Massardi \& Burigana 2010) was applied to predict when our target sources were being observed by the satellite, according to its publicly available pre-programmed pointing list\footnote{Informations on the {\it Planck} scanning strategy and pointing are publicly available for external observers at \url{http://www.sciops.esa.int/index.php?project=PLANCK\&page= Pointing}.}.

The SiMPlE project obtained 21 epochs of allocated time in the period between June-December 2010. Each epoch was scheduled for up to 24 hours. Only one epoch was completely lost because of adverse weather conditions, which also affect 8.3 and 5 GHz observations.
In the first six months of the SiMPlE project no 22-GHz acquisitions could be performed because of the insufficient sensitivity of the available K-band receiver. The temporary installation - for commissioning purposes - of a multi-feed 18-26 GHz receiver, whose final destination is the new Sardinia Radio Telescope, allowed us to carry out 22-GHz observations, together with new lower frequency ones, during the first semester of 2011.

The scheduling procedure prioritised for each epoch the sources that were being observed by {\it Planck}, then the sources that in previous runs had been flagged out because of poor data quality, and, finally, all the other sources, in order to observe the sources as many times as possible.

\subsection{OTF cross scans and scheduling criteria at the Medicina radio telescope}

Observations were carried out with the 32-m single dish of the Medicina radio telescope in the OTF scan mode (Mangum et al. 2007), exploiting both hardware devices and software tools recently developed, at present still under commissioning.
On the hardware side, a new analogue backend was employed, the first one to be fully dedicated to continuum single-dish activities with the Medicina VLBI antenna. 
Besides permitting the execution of high-speed scans, is very helpful to avoid system instabilities and to better trace atmospheric variations.
Fast OTF scans were made possible as the antenna was provided with a new control system, the Enhanced Single-dish Control System (ESCS)  specifically designed to perform single-dish observations exploiting the full potential of the telescope.
Tests have demonstrated that scans can be carried out up to $20^{\circ}$/min without compromising the pointing accuracy. 
The optimised setup for the cross-scans required acquisitions at much lower speeds, with a sampling rate of 25 Hz and a resulting spatial resolution of 60 samples/beam. Table \ref{tab:dettos} lists the main scan setup parameters employed.

\begin{table}
\caption{Observational details used to plan the cross scans; in each scan the time on-source is of about 2.5 s.
The high system temperature of the 8.3 GHz receiver was due to malfunctions in the receiver cryogenic system.
}
\begin{tabular}{ccccccc}
\hline
\tiny{Frequency} & \tiny{Beam} & \tiny{Scan }  & \tiny{Scan }     & \tiny{Usable}    & \tiny{Tsys} & \tiny{Instant} \\
                 & \tiny{size} & \tiny{length} & \tiny{speed}     & \tiny{bandwidth} & \tiny{}     & \tiny{rms}     \\
\tiny{[GHz]}     & \tiny{[$'$]}& \tiny{[HPBW]} & \tiny{[$'$/min]} & \tiny{[MHz]}     & \tiny{[K]}  & \tiny{[mJy]}   \\
\hline
5.0   &  7.5   &  5 &   180  &  2$\times$ 80   & 30  &   74.1 \\
8.3   &  4.8   &  5 &   120  &  2$\times$ 230  & 80  &  132.7 \\
\hline
\end{tabular}
\label{tab:dettos}
\end{table}

The number of cross-scans to be performed on each source was adjusted in real time during the observations, according to the actual system performance and weather conditions, in order to reach a SNR of at least 10 in the final integrated scans.
This dynamical scheduling was possible thanks to an on-purpose-developed tool: the Positional On-the-flight-scan Planner (POP).
As run-time execution of this tool completes in few seconds, POP allowed us to quickly produce new schedules whenever the weather and/or system conditions changed during an observing session. The only input required to the observer consists in the target coordinate list, the LST time at which the schedule is supposed to be executed, the observing frequency and, optionally, the estimated flux density of the sources.
POP schedules the sources after checking for their visibility, taking into account the telescope movement limits and duration in azimuth and elevation, sorting them in ascending RA, to minimise the slewing time between targets and maximise the number of observed targets during the allocated time.

For each source POP calculates the minimum number of scans required to reach a noise level 10 times smaller than the 23-GHz WMAP flux density, provided in the input position list. If no flux density is provided POP schedules for each target a number of scans equal to a user-defined value indicated in the software configuration settings: this is of use in case of bad weather conditions or instrumental failures when a longer integration time is needed.

Table \ref{tab:scans} shows some examples of the cross scans required to obtain SNR=10 for different flux density limits.

Each sub-scan on the sample sources is 5 Half Power Beam Width (HPBW) long, both at 5 and 8.3 GHz, in order to always have enough off-source samples and thus to better evaluate the background. In order to keep the sampling rate fixed at different frequencies (60 samples/beam), it required to change the scan speed, according to the different HPBW. This means the Gaussian produced by the source is characterised by an high sampling rate, allowing to perform the fit on many more points. Scans on calibrators were set larger (7 HPBW): because of their crucial role in the sources flux densities recovery, the larger number of off-source samples provides a more precise identification of the baseline, translating into a more reliable measure for the parameters which will subsequently be employed in the calibration phase (details in Section \ref{sec:datared}).

\begin{table}
\caption{Number of cross scans performed to achieve SNR=10 on sources of the given flux density limit, considering average weather conditions and the actual system setup exploited during the observing session. A minimum of 2 cross scans was carried out even on the brightest sources, despite a single scan would have been more than sufficient from the sensitivity point of view, to cross-check for possible artefacts in the data.
}
\begin{tabular}{ccc}
\hline
\tiny{Flux density limit} & \tiny{5 GHz} & \tiny{8.3 GHz}  \\
\tiny{[mJy]}& \tiny{cross scans} & \tiny{cross scans}   \\
\hline
200   &  7   &  22  \\
500   &  2   &  4   \\
1000  &  1   &  1   \\
\hline
\end{tabular}
\label{tab:scans}
\end{table}

\subsection{Region mapping}
\label{sec:maps}

Some sources in the NEWPS five-years $5\,\sigma$ catalogue are classified as {\it undefined} because of the lack of counterpart in low frequency catalogues. Some other objects have multiple associations within the beam size with low frequency wide area surveys (NVSS and/or GB6). These samples might include spurious detections on WMAP maps as well as extended sources or peaks of Galactic foreground emission erroneously identified as sources by the detection procedures.

In order to identify or to resolve the candidate sources we scanned patches of the sky around the source positions. We selected a sample of 29 undefined targets using coordinates of the most likely counterpart in the GB6 or NVSS catalogues when possible, otherwise we used the coordinates of the NEWPS detection. \\
For each of these sources, we pointed and mapped patches in the sky of about $50' \times 50'$, centred in the estimated position of the sources. The scanning strategy consists of 21 equally spaced scans in each direction (RA-DEC). Furthermore, consecutive scans are performed with opposite movement direction. To get a visual display of the mapped regions, we have built a grid based on the RA and DEC coordinates of the considered source and we associate the average value of a scan lying in each grid step to the corresponding RA, DEC position. The resolution of the grid on which the map is built could be changed, but in general, for a good visualisation of the sky patch, we decided for values ranging from 20 to 25 grid steps. This is practically equivalent to a re-bin of the scans in steps ranging from $1.75$ to $1.4$ arcmin.

\section{Data reduction}\label{sec:datared}

\subsection{Data quality inspection and flagging}

In optimal conditions for observations, each scan has the shape of the beam transfer function, i.e. a Gaussian with a FWHM corresponding to the beam size\footnote{ The measured deviations of the beam shape from a circular Gaussian beam are less than 2 per cent, including the measure errors.}, overlapped to a baseline that corresponds to the off-source zero level of the signal. Along the scans, the fluctuations of amplitude are given by a Gaussian noise.
However, cloudy weather, the presence of random contributions by radio frequency interference (RFI) or digital noise heavily affected portion of the data appearing as bumpy baselines or spike-like features.

\begin{figure}
 \centering
 \includegraphics[width=.95\hsize]{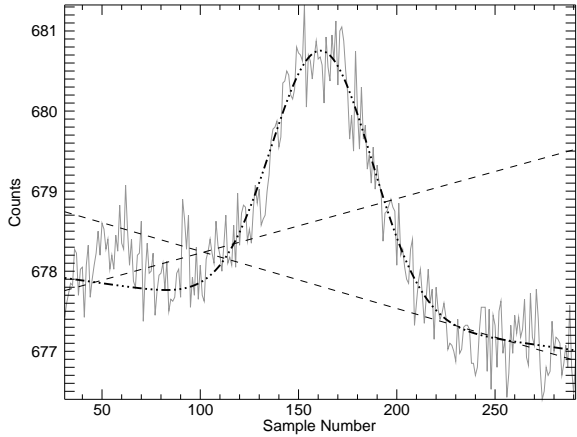}
 \caption{A sub-scan of a source (solid grey line) displayed by the flagging procedure. Here are also plotted the Gaussian fit of the scan (3 dot-dash line) and the linear fit of the left-right-hand baseline (dashed lines).
 }
\label{fig:flag}
\end{figure}

Hence, by fitting a Gaussian function and a linear baseline on the data it might be possible to assess the quality of each scan
comparing the fitted value with the expected values and the overall goodness of the fit trough a $\chi^2$ analysis. A similar test allows one to draw considerations about the Gaussian FWHM, the length of the baselines, the signal to noise ratio, the differences between the slopes of the two parts of the baseline (i.e. left and right with respect to the Gaussian).
The broad variety of possible scan behaviours and the large number of parameters to be considered did not allowed to easily limit the parameter space. Furthermore, in order to characterise the instrumental capabilities it is interesting, during the on-going commissioning phase, to classify the problems that affect the data.
For this reasons we decided to manually inspect the whole dataset to classify the scans according to their quality and to the features that affect them, if any is present. Hence, we have developed a tool that performs the fit of a Gaussian plus a linear baseline, showing it overlaid to the scan data as a reference (see Fig. \ref{fig:flag}) in the visual inspection to guarantee homogeneity to the flagging criteria.

We pay particular attention to the contributions affecting the linearity of the baseline: bumps, RFI, different slope of the left and right baseline arms. These features heavily affect the linear and the Gaussian fit of each scan and thus scans presenting any of these contributions are rejected. Of course any scan characterised by an odd shape of the beam transfer function is rejected as well.

Bad weather conditions also reduce the SNR, so that in the worst cases (i.e. faint sources with high noise level) the sources were completely embedded in the noise fluctuations. Averaging over several scans reduces the noise and amplifies the SNR, thus for these critical sources the flagging was performed also on the integrated scans.

In future phases we plan to automatise the flagging process on the basis of a statistical analysis of the flagging done so far, to minimise the subjectivity introduced in the flagging procedure by visual inspection.

\subsection{OTF scan processing}
\label{sec:calib}

We have developed the OTF Scan Calibration-Reduction (OSCaR) pipeline for the data reduction of calibrators and target sources. It constitutes an easily customisable ensemble of IDL routines capable of handling up to huge amounts of data, operating at all the SiMPlE frequencies.
 
A run of the entire pipeline does not require a large amount of memory as well as a huge computational power: for about ten days of flagged data the complete process requires less than ten minutes\footnote{On a machine equipped with a 2.0 GHz Intel Core 2 Duo and 3 Gigabytes of RAM memory.}. The entire software package can be easily modified to be used for projects involving OTF cross-scan observations. 

\begin{figure}
 \centering
 \includegraphics[width=1.\hsize]{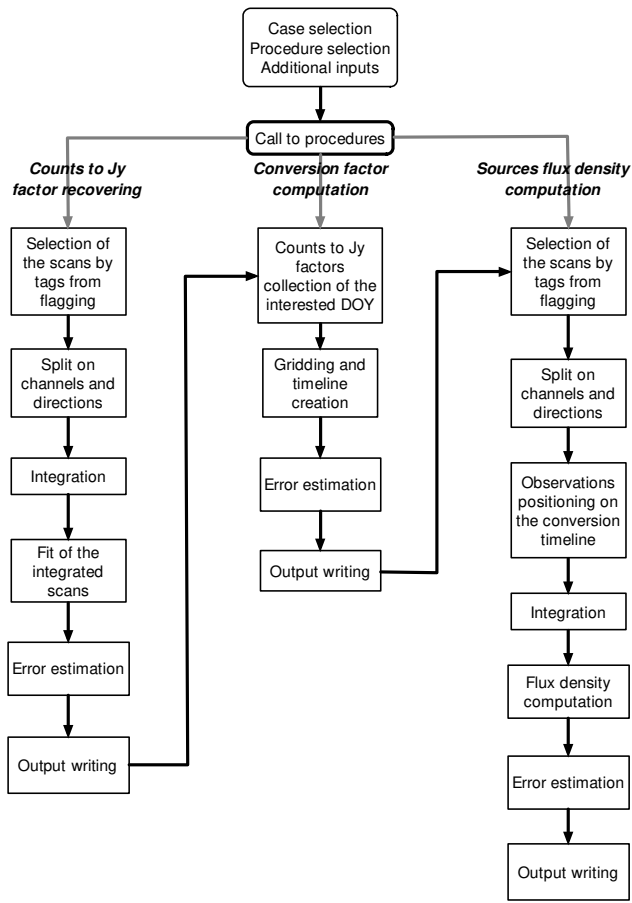}
 \caption{Scheme of the main operations in OSCaR. See \ref{sec:calib} for a detailed description of the procedures.
 }
\label{fig:oscar}
\end{figure}

The main steps of OSCaR (see Fig. \ref{fig:oscar}) are described in the following sections and can be summarised as follows: {\it i}) the estimation of the factor to convert the count sampling levels to Jy units by rescaling the counts measured on calibrators to their known flux densities; {\it ii}) the reconstruction of the variation with time of the counts-to-Jy factor and of the component of the flux density error due to calibration; {\it iii}) the calculation of the source flux densities and their errors, by applying the correct counts-to-Jy factor, integrating the scans for a source, and fitting them with a Gaussian that reproduces the receiver response function. 

\subsubsection{Recovering of the counts to Jy conversion factor.}

The first task of OSCaR consists in recovering the conversion factors to transform the raw signal intensity, given in arbitrary counts, to a calibrated flux density (Jy). This is done by fitting the scans with a Gaussian fit, plus a linear fit, according to the following:
\begin{equation}
f(x) = A_0\; {\rm exp}(-z^2/2)+ A_3 + A_4x\;,
\end{equation}
with
\begin{equation}
z = \frac{x - A_1}{A_2}\;.
\end{equation}

In these equations, $A_0$ is the height of the Gaussian, $A_1$ is the centre of the Gaussian, $A_2$ is the width of the Gaussian, $A_3$ is the constant normalisation term, and $A_4$ is the slope of the baseline. In this way, we measure the Gaussian curve amplitude and compare it with the source absolute flux density. The flux density/amplitude ratio is the counts-to-Jy factor valid for the observed elevation position.

We consider as primary calibrators 3C286 and 3C295, the only two sources reporting negligible variability by Ott et al. (1993) with respect to the original scale of Baars et al. (1977). This flux density stability has been confirmed by recent Effelsberg observations (Kraus \& Bach, priv. comm.).
Then, whenever it is possible, we measure the flux density of other calibrators (3C147, 3C48, 3C123, NGC7027, DR21) against the primary ones, exploiting the counts-to-Jy factors recovered on the primary calibrators.  This allowed us the possibility of at least one calibrator observation per time interval of few hours (typically 2-3 hours in good observing conditions). 

\begin{figure}
 \centering
 \includegraphics[width=1.\hsize]{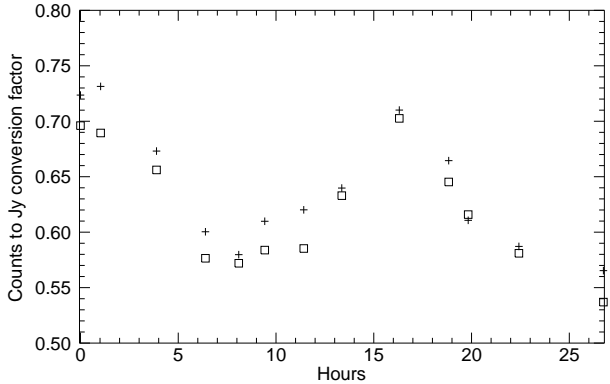}
 \caption{Conversion factor timeline recovered from an entire observing session. The cross and square symbols represents left and right channel respectively. 
 }
\label{fig:conv_fact}
\end{figure}

A timeline of the conversion factors is recovered along all the days of acquisitions.
In particular, once we collected all the conversion factors on an entire session of observation (see Fig. \ref{fig:conv_fact}) we interpolated them in time, obtaining a continuous projection of these factors in the interval between two contiguous calibrator observations.  
We chose this strategy in order to have a time dependent conversion factor, reflecting in a smoother way the oscillations of each computed factor during the day. These factors vary whenever a change happens in the weather conditions - implying a different atmospheric absorption affecting the source flux density - and when the system gain, for intrinsic reasons or user-defined choices,  changes as well. Higher frequency observations are more sensitive to these contributions, and thus need a more frequent and accurate evaluation of the counts-to-Jy factors.

\subsubsection{Flux densities.}

The flux density measurements on the observed sources are performed through an algorithm which is very similar to the one used in recovering the counts-to-Jy conversion factors. The equation for the Gaussian fit is also the same.
The accepted scans are selected and integrated according to the tags provided during the flagging phase. The integration is performed through a precise positional alignment, in turn allowed by the very reliable antenna pointing system. It is also possible to define the time interval within which more scans on the same source must be considered consecutive and thus integrated, otherwise they are considered as separated measurements, each processed with the proper calibration factor. 

Before proceeding with the flux density computation a further check on the HPBW of the scan is performed. We accepted scans with values of the HPBW within 20 per cent variation from the nominal value. This criterion selects not only the scans for which the instrument behaved properly, but can also remove from our analysis sources which are extended by more than 1.2$\times$HPBW, for which our flux density estimates are only a lower limit.

We obtain separate flux density estimates for the two acquisition channels and for each scan direction. We firstly compute the flux density from the cross created by the integration of the RA-DEC scans in each channel, then through a weighted average we achieve a unique flux density for the considered source. The application of the counts-to-Jy factor takes into account that the source and the calibrator had been observed at different elevations. The conversion factor is properly rescaled by means of the standard gain-elevation curve provided for the Medicina dish. The equations are the following:
\begin{equation}
G = (- 4.6834953\cdot 10^{-5} \cdot e^2) +
\end{equation}
$$
+ (6.2403816\cdot 10^{-3} \cdot e)+7.9212981\cdot 10^{-1}\,,
$$
at 5 GHz, while at 8.3 GHz it becomes
\begin{equation}
G = (- 7.2457279\cdot 10^{-5} \cdot e^2) +
\end{equation}
$$
+ (1.0623634\cdot 10^{-2} \cdot e) + 6.1059261\cdot 10^{-1}\;.
$$
$G$ represents the elevation dependent normalised gain, while $e$ is the elevation of the considered source/calibrator.
A flux density compensation related to a possible offset in the pointing of the sources is also taken into account according to the following equation:

\begin{equation}
\Delta S^{offset}_{j,k} = exp[-(x \cdot 1.66/(HPBW \cdot \pi))]
\end{equation}

\begin{figure}
 \centering
 \includegraphics[width=1.\hsize]{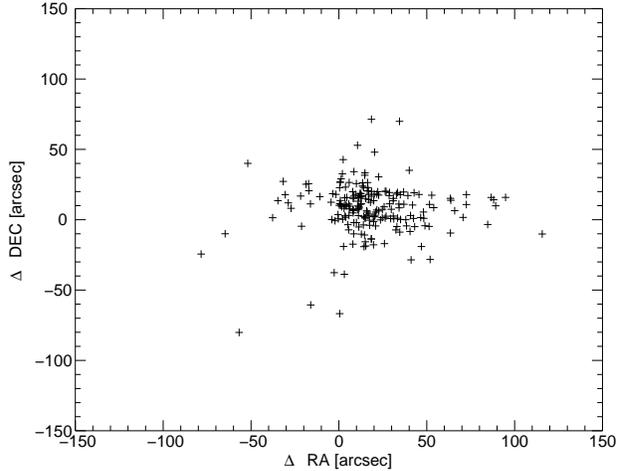}
 \caption{The difference between the input and the recovered RA and DEC position of the Northern Sample sources for about 200 individual pointings at 5GHz (for comparison the beam FWHM is 7.5 arcmin).
 }
\label{fig:displ}
\end{figure}

In this equation $j=RA,DEC$, $k=L, R$ indicating left and right channel, $x$ represents the positional offset according to the direction considered, and $HPBW$ is the nominal value of the beam at the frequency considered. We calculate the positional offset by comparing the position of the peak detection performed by the fitting function with the input coordinates values. The positional offset of about 200 individual pointings is shown in Fig. \ref{fig:displ}.

Finally, each source is calibrated using a conversion factor recovered matching the acquisition epoch to the calibration timeline.
In this way we ensure a short time distance between the observed source and the calibration factor, in order to minimise the possibility of weather or system variations.\\

The set-up of the run is interactively performed after launching the main program. Among the available options, there is the possibility to call each procedure independently, to provide a list of sources to be analysed or to work on an entire observing session.

Through the tags provided during the flagging phase we analyse each channel of each sub-scan independently. While this procedure would be standard for a polarisation analysis, for the total intensity study we have the possibility to choose to use one or both channels of each sub-scan.\\
Particular attention was dedicated to the handling of the time tag of each scan. We aimed at collecting as many observations as possible, of each source, at different times. These time intervals can span from hours to weeks or months. Thus we implemented the pipeline in order to have a coherent integration of the scans concatenated in time. The automatisation of this strategy of integration permits to easily handle multiple observations of calibrators in a single day and consequently the creation of the detailed calibration timeline we use to calculate the  flux densities. The possibility to recover the flux densities of selected sources in more observing sessions at once makes OSCaR well suited to evaluate variability as well.

We could have relied on the telescope pointing/acquisition system for an immediate, sample-by-sample integration of the scans. Through a dedicated test, we confirmed that the maximum misalignment encountered, comparing the initial positions of the scans, corresponds to one sample of data (7.2 arcsec for $3^{\circ}$/min scans, i.e. less than 2 per cent of the HPBW). However, for a more precise estimation of counts-to-Jy conversion factors and flux densities, it is important to avoid any possible offset, thus any misalignment, even of such entity, is corrected during the integration of the scans.

\subsubsection{Flux density error estimation}

The error budget in the final flux density results from two main components: the calibration error $S_{cal}$ and the noise error $S_N$

\begin{equation}
	\Delta S_{TOT}=\sqrt{\Delta S^{2}_{cal}+ \Delta S^{2}_{N}}\,.
	            \label{eq:DELTAS-TOT}
\end{equation}

The two terms have identical physical causes, although with different weights, and can be treated using a common formalism.
The contribution to each component is given by:

\begin{equation}
\Delta S_{i}=\sqrt{\delta^{2}_{baseline}+ \delta^{2}_{Gauss}}\;,
\label{eq:DELTAS-i}
\end{equation}

with $i\,=\,cal,N$.

The baseline error ($\delta_{baseline}$) is mainly due to the white instrumental and background noise and to the varying influence of the ground radiation while the antenna is scanning at different angles. It has two spectral components: one showing white spectrum (background, atmosphere, statistic uncertainty) and the other characterised by 1/f behaviour (long time drift of detectors and slow atmospheric changes between the telescope and the target). The confusion limit, the asymptotic limit of the white component in the case of a large number of samples, represents only a part of the baseline error.

In order to compute this contribution we subtract the linear fit, performed on the scan neglecting the Gaussian bell, from the baseline. 
This reduces the long time drift effect. Then we calculate the standard deviation of the subtracted data. In this way, we take into account both the white noise contribution and the spurious residual effects on time scale of a single-source observation. As a general rule, we decided to keep margins, when fitting the linear baselines, neglecting the first ten points on the left-hand and on the right-hand sides of the Gauss fit, whose width was calculated from the beam size.

The Gauss error ($\delta_{gauss}$) estimates the goodness of the Gaussian fit. It represents the error committed in calculating the coefficient $A_{0}$ (amplitude of the Gaussian) of the exponential term, and can be ascribed either to white scatter effects affecting the whole scan or to local drops and spikes of signal, mostly  in correspondence of the tip. However, in most cases, the preliminary operation of data flagging is expected to have already prevented the Gaussian shape from evident anomalies.
Operatively, the Gauss fit is calculated over the integrated scans and the associated error is provided by the fitting routine itself. The final error in the flux density, indicated in Eq.~(\ref{eq:DELTAS-i}), comes out as the amplitude uncertainty of the height of the bell estimated through the Gauss fit.

Because of the flagging strategy adopted, acting individually on each subscan, the number of good data was different for RA and DEC scans as well as for the two polarisation channels, for each source observed. For this reason, Eq.~(\ref{eq:DELTAS-i}) was calculated also for right-hand channel (RHC) and left-hand channel (LHC) when scanning in RA and in DEC.

In accordance with the observing strategy, about 20 sources were observed in the time window between two calibrators\footnote{ Performing an OTF cross-scan on a source takes approximatively 30 sec, including pre and post scan operations.}. In a few cases the primary calibrators were not available during the whole observing session: consequently, we have been forced to calculate the calibration constants using secondary calibrators. Through dedicated tests, we found out that this strategy could affect the calibration error by less than 5 per cent in a day with good observing conditions. Therefore, we decided to calculate a weighted calibration error for each source, relating the calibration error to the time window considered instead of calculating an average error over the whole run.\\ 
The calibration error on each observed source comes out from weighting the errors committed on the two nearest calibrators by the distance in time of each source from the calibrators. Hence, given $\tau_{1}$ and $\tau_{2}$ the distance in time of the i-th source from calibrators 1 and 2 (characterised respectively by errors $\Delta S_{cal_1}$and  $\Delta S_{cal_2}$), the weighted error is:

\begin{equation}
\Sigma_{CAL}=\tau_{2}\frac{\Delta\,S_{cal_1}}{\tau_{1}+\tau_{2}}+\tau_{1}\frac{\Delta,S_{cal_2}}{\tau_{1}+\tau_{2}}
	            \label{eq:sigma-cal}
\end{equation}

Similar to $\Delta S_{i}$, $\Sigma_{CAL}$ was calculated for channels RHC and LHC both in RA and DEC; when combining these error terms with  Eq.~(\ref{eq:DELTAS-TOT}) we get the following four-term array:
\begin{eqnarray}
\Delta S_{RA_{L}} =\sqrt{\Sigma_{CAL^{RA_{L}}}^{2}+ \Delta S_{N^{RA_{L}}}^{2}}\\
\Delta S_{RA_{R}} =\sqrt{\Sigma_{CAL^{RA_{R}}}^{2}+ \Delta S_{N^{RA_{R}}}^{2}}\\
\Delta S_{DEC_{L}} =\sqrt{\Sigma_{CAL^{DEC_{L}}}^{2}+ \Delta S_{N^{DEC_{L}}}^{2}}\\
\Delta S_{DEC_{R}} =\sqrt{\Sigma_{CAL^{DEC_{R}}}^{2}+ \Delta S_{N^{DEC_{R}}}^{2}}
			 \label{eq:DELTAS-ARR}
\end{eqnarray}

\begin{figure}
 \centering
 \includegraphics[width=.95\hsize]{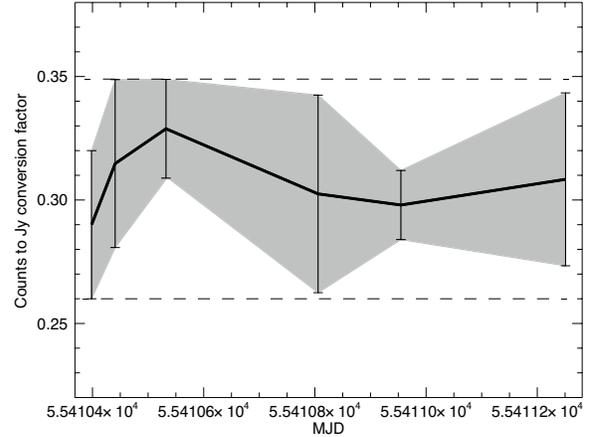}
 \caption{A typical conversion factor timeline. The black solid line represent the effective conversion factor applied to the data, while the grey zone represents the error associated with the factors themselves. The dashed lines represents a rough estimate of the error that can be considered for the whole day of year (DOY).
 }
\label{fig:err_c}
\end{figure}

The calibration errors, calculated following the above procedure, were hence compared with the error calculated just considering the whole error bar obtained when displaying the calibration constants referring to all the calibrators available during the observing day considered. This comparison is displayed in Fig. \ref{fig:err_c}. 
\\
The final error $\Sigma_{FX}$ in the flux density of each source comes out as the square root of the sum of the squares (RSS).\\

\begin{equation}	
\Sigma_{FX}=\sqrt{	\Delta S_{RA_{L}}^{2}+	\Delta S_{RA_{R}}^{2}+\Delta S_{DEC_{L}}^{2}+	\Delta S_{DEC_{R}}^{2}}
	            \label{eq:SIGMA-FX}
	\end{equation}

\begin{figure}
 \centering
 \includegraphics[width=1.\hsize]{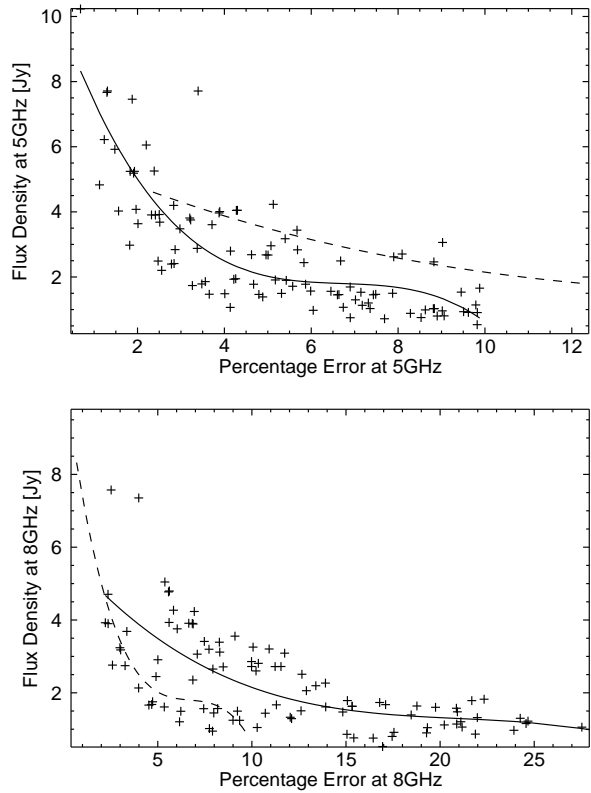}
 \caption{ The flux densities from 100 measurements of the Northern Sample sources (crosses) versus their percentage error are plotted. The upper panel shows the 5-GHz measurements, while the lower those at 8.3 GHz. The black lines represent the fits of the plotted data and the same quantity for the other frequency is overplotted for comparison (dashed lines). Multiple observations of the same sources are also considered.
 }
\label{fig:errori}
\end{figure}

The error contribution can be mapped through a plot, showing how it varies (in percentage) with the source intensity (Fig. \ref{fig:errori}), at 5 and 8.3 GHz.

\begin{figure*}
\centering
\includegraphics[width=1.\hsize]{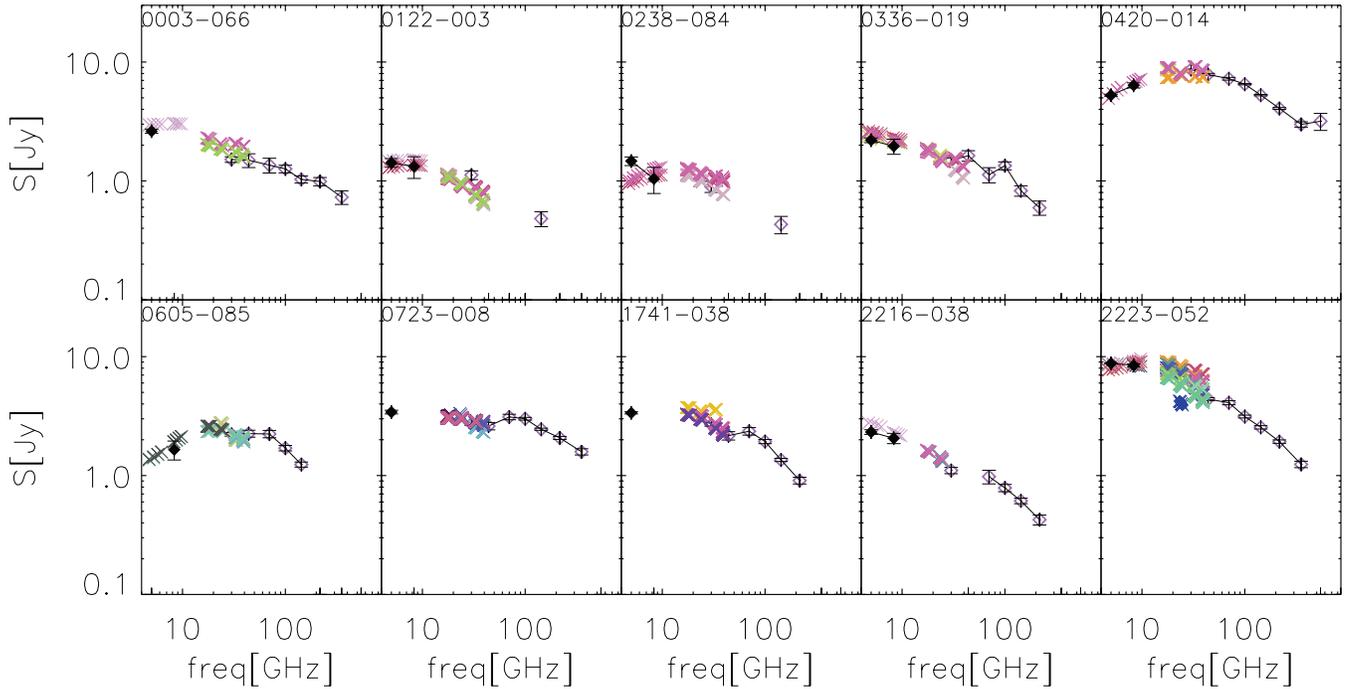}
\caption{SEDs for a sample of equatorial targets observed simultaneously with the Medicina radio telescope (filled diamonds), the ATCA (cross, in the framework of the PACO project) and the {\it Planck} satellite (empty diamonds, data from ERCSC, The {\it Planck} Collaboration 2011a). The target 0238-084 was observed from Medicina with some days of delay. The displacement in its SED could be due to variability of the source itself.  
}
\label{fig:PACO_SiMPlE}
\end{figure*}

\subsection{Simultaneous ATCA-Medicina observations}
In addition to the NEWPS Northern hemisphere sample, we have observed with the Medicina radio telescope a list of 12 sources with $\delta<0^\circ$ among the targets of the PACO project. We observed almost all these targets simultaneously with the two facilities. In any case, the observations were performed within 10 days from the {\it Planck} satellite ones. The PACO project observed them in 24$\times$512~MHz--wide frequency sub-bands between 4.5 and 40 GHz, including data at 5.244 and 8.232 GHz.
This sample, although small, has been used as a test to verify that our procedures give consistent results. The two projects in fact are characterised by the use of different telescopes, data-reduction pipelines, and completely different calibration schemes and sources.

The SEDs of the equatorial targets including the Medicina and ATCA radio telescopes, and the {\it Planck} satellite data are in Fig. \ref{fig:PACO_SiMPlE}, showing a fairly good agreement.

\begin{figure}
\centering
\includegraphics[width=1.\hsize]{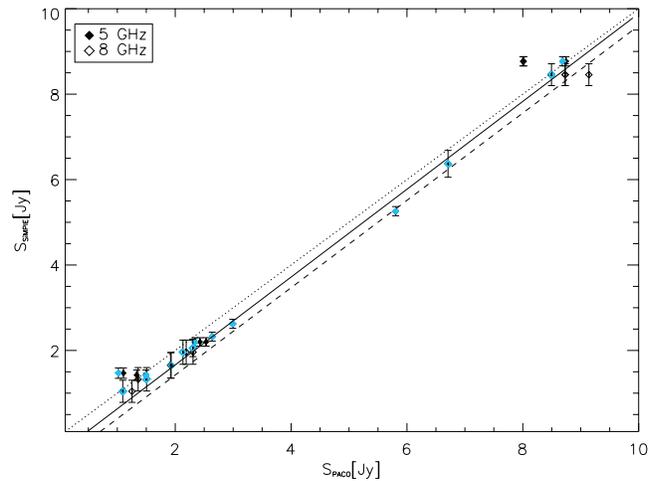}
\caption{Flux density comparison between SiMPlE and PACO data at 5 and 5.244 (filled symbols) and at 8.3 and 8.232 GHz (empty symbols) respectively. No corrections for the small difference in frequency have been applied. The best fitting lines at 5 and 8 GHz are respectively the solid and dashed lines. The dotted line represents unity. 
}
\label{fig:SS_PACO_SiMPlE}
\end{figure}

The comparison is unbiased by source variability and only depends on the instrumental properties. The {\it Planck} satellite lower frequency channel is too high to transfer its CMB dipole based flux density calibration to the low-frequency Medicina telescope channels, but we can compare the two ground-based facilities flux density scales. The flux density calibration for the ATCA data is based on the observation of one single very stable source, PKS~B1934-638.
Fig. \ref{fig:SS_PACO_SiMPlE} shows the flux density comparisons at 5 and 8.3 GHz.
The best fitting line at 5 GHz has a slope equal to $1.03\pm0.09$ and crosses the y-axis at $(-0.39\pm0.36)$ Jy. At 8 GHz it has slope equal to $1.02\pm0.15$ and crosses the y-axis at $(-0.62\pm0.65)$ Jy: the flux density is consistent within the error bars between the two instruments. A small bias in calibration is probably present, but the small sample does not allow to quantify it.
Further observations of a larger equatorial sources sample are on-going to quantify the calibration differences between the telescopes with smaller uncertainties and to have a common calibration scale between several facilities and including at mm frequencies also the {\it Planck} satellite.

\section{The $\delta \geq 45^\circ$ sample}\label{sec:catalogue}

After the flagging and the removal of data points with signal-to-noise ratio lower than 5,  67 of the 69 NEWPS selected sources with $\delta \geq 45^{\circ}$ observed with OTF cross-scans have reliable data at 5 GHz (the exceptions are the 179th and the 245th in the NEWPS catalogue for which all the data were flagged for bad weather conditions) and 59 sources have valid data also at 8 GHz in at least one epoch in the June-December runs. Table \ref{tab:catalogue} lists the data at 5 and 8.3 GHz for one observing epoch for each source per row. The columns are as follows:
\begin{description}
\item[1] Source identification. For sources in the NEWPS catalogue the id corresponds to the sequential number in the Massardi et al. (2009) catalogue.
\item[2-3] Right ascension in hour and declination in degrees.
\item[4] Epoch of observation in YYYY-MM-DD.
\item[5] Flag asterisks for simultaneity with the {\it Planck} observations.
\item[6-7] Flux density at 5 and 8.3 GHz in mJy.
\item[8-9] Flux density error at 5 and 8.3 GHz in mJy.
\end{description}

10 more targets have been observed by mapping the surrounding sky region and will be discussed in the following section.

\subsection{Region maps with $\delta \geq 45^\circ$}
\label{sec:mapping_dataanalysis}

\begin{figure*}
 \centering
 \includegraphics[width=.8\hsize]{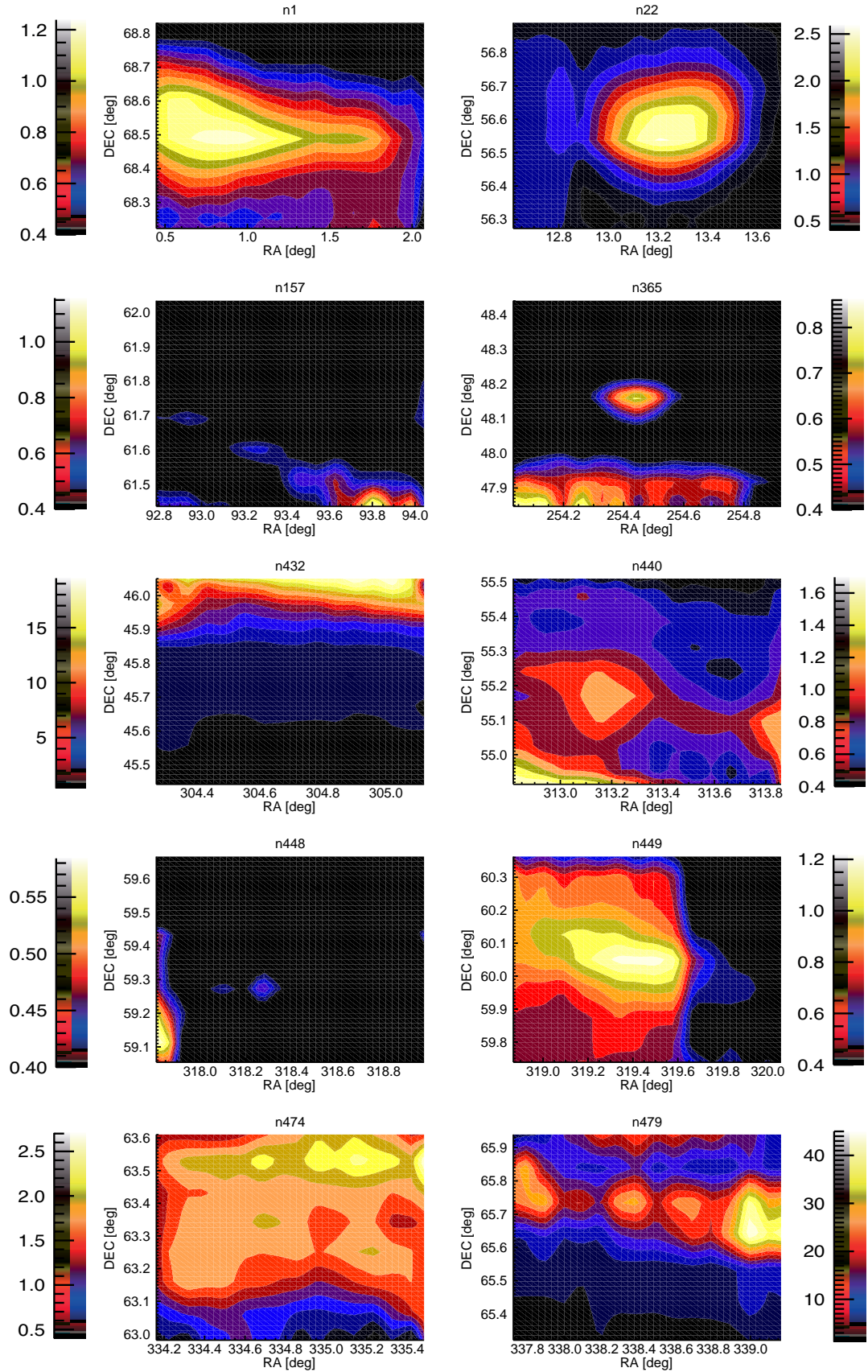}
 \caption{The ten regions of sky mapped around the undefined sources described in \ref{sec:maps}. The flux density scale is expressed in Jy.
 }
\label{fig:all_maps}
\end{figure*}

Among the regions that we have mapped at 5 GHz, ten have declination $\delta \geq 45^\circ$. 
The maps are shown in Fig. \ref{fig:all_maps}.

For each map we perform a two-dimensional Gaussian fit, in order to identify the candidate source and estimate a lower limit to the source flux densities. 
This procedure fits an elliptical Gaussian equation to gridded data. The fitting function is:
\begin{equation}
F(x,y) = A_0 + A_1 \cdot {\rm exp}(-U/2)\;,
\end{equation}
with the elliptical function described by $U = (x' /a)^2 + (y'/b)^2$, while $A_0$ and $A_1$ are the constant term and the scale factor, respectively. The rotated coordinate system is defined as:
\begin{eqnarray}
x' = (x - h) \cos(T) - (y - k) \sin(T)\\
y' = (x - h) \sin(T) + (y - k) \cos(T)
\end{eqnarray}
In our case, $x$ and $y$ are the scan directions RA and DEC, $(h, k)$ is the centre of the ellipse, and $T$ is the rotation angle from the $x$ axis in clockwise direction. Furthermore, it is possible to recover the width of the Gaussian in the two directions, the location of the centre on the two axes, and the rotation of the ellipse with respect to the $x$ axis.
Once the peak position has been identified we estimated the peak flux density which is a lower limit to the source flux density in case a candidate could be identified. As the profile of an extended source could be far from a Gaussian this approach has been used only to identify a peak and estimate a peak flux density. As the peak flux density is only a lower limit of the integrated flux density, we define a candidate source detection only the case in which the peak flux density is at least 3 times larger than the noise estimated on the map and the source can be univocally identified in the map (i.e. there is only one peak and the scans passing across it  passed the visual flagging procedure).
Table \ref{tab:mappes} lists the fit details for the 10 maps in the Northern sample region. 

\begin{figure}
 \centering
 \includegraphics[width=1.\hsize]{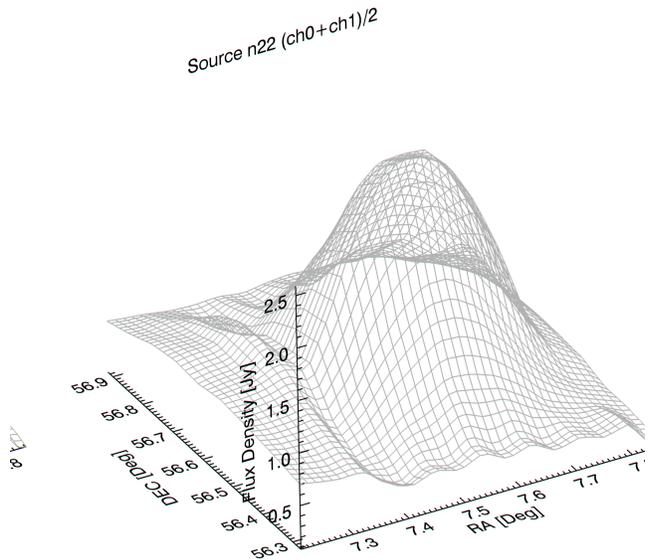}
 \caption{Surface projection of the sky region centred on NEWPS source n22 identified with the HII region NGC~0281.
 }
\label{fig:n22}
\end{figure}

5 of the 10 mapped regions with $\delta \geq 45^\circ$ have been identified as candidate sources, but only three of them have (peak) signal-to-noise ratio larger than 5. Fig. \ref{fig:n22} shows a graphical elaboration of the map of n22, one of the five source detections. This object has been identified with the HII region NGC~0281. The recovered peak position of the 5 detected sources was then compared with that provided by WMAP: in all the cases the displacement does not exceed 10 arcmin.

Hence among the 79 sources identified in the NEWPS sample with $\delta \geq 45^\circ$, about $95$ per cent have been confirmed as genuine/candidate sources. A proper analysis of the comparison with the selection flux densities will be done in future works where we will present the 22-GHz data for our objects, for which the observing campaign is on-going.

\begin{table}
\caption{Peak flux densities extracted for the 10 mapped regions with $\delta \geq 45^\circ$. The last column indicates as 1 the sources that have $S_{peak\,5GHz}/\sigma>3$ and the scans across the peak position passed the visual flagging procedure.}
\label{tab:mappes}
\begin{tabular}{lccccc}
\hline
\tiny{NEWPS} & RA & $\delta$ & $S_{peak\,5GHz}$& $\sigma$& candidate\\
\tiny{       ID} & [deg] & [deg] &  [Jy] &  [Jy]& flag\\
\hline
n1  &   00:03:38& 68:28:50&  1.24& 0.11& 1\\
n22 &  00:52:56& 56:35:22&  2.59& 0.26& 1\\
n157&  06:14:21& 61:28:32&  1.17& 0.35& 0\\
n365& 16:57:46& 48:08:32&  0.71& 0.35& 1\\
n432& 20:19:36& 46:03:05& 21.82& 0.34& 0\\
n440& 20:52:47& 55:11:44&  2.05& 0.78& 0\\
n448& 21:13:23& 59:21:51&  0.79& 0.32& 0\\
n449& 21:17:36& 60:02:45&  1.22& 0.36& 1\\
n474& 22:19:56& 63:33:41&  3.08& 0.14& 0\\
n479& 22:36:05& 65:44:21& 65.66& 0.46& 1\\
\hline
\end{tabular}
\end{table}

\subsection{Spectral behaviour and comparisons with other catalogues}

\begin{figure}
 \centering
 \includegraphics[width=1.\hsize]{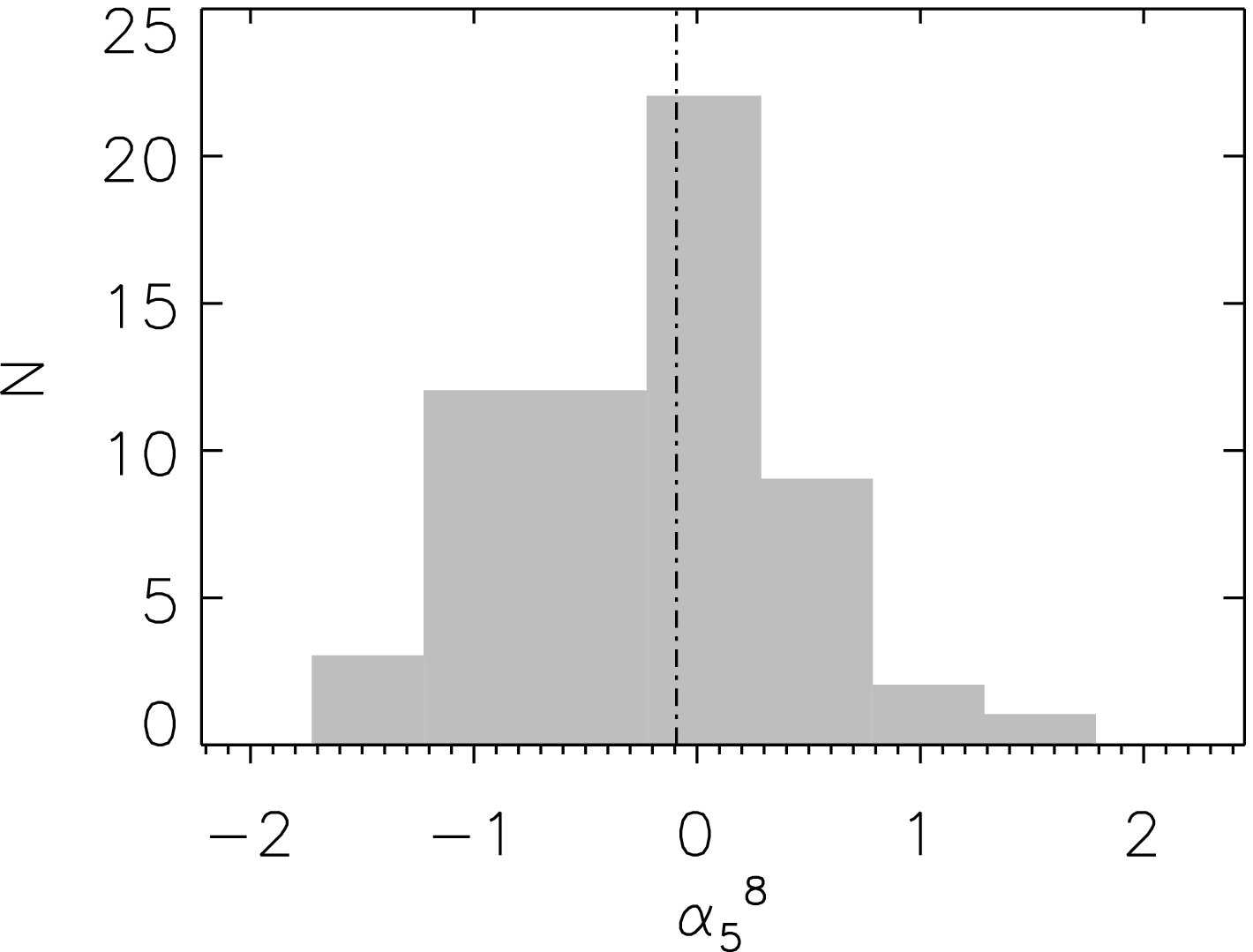}
 \caption{Distribution of spectral indices in the 5-8 GHz frequency ranges.
 }
\label{fig:alpha_5_8}
\end{figure}
\begin{figure}
 \centering
 \includegraphics[width=1.\hsize]{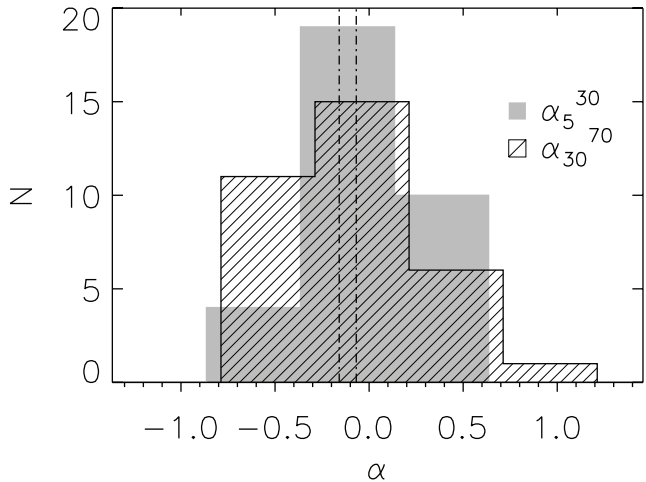}
 \caption{Distribution of spectral indices in the 5-30 (shaded area) and 30-70 GHz (hatched area) frequency ranges. The bin size is fixed at 0.5 for both the histograms. The displacement is due to the different ranges characterising the two spectral indexes.  
 }
\label{fig:alpha_30_70}
\end{figure}
\begin{figure}
 \centering
 \includegraphics[width=1.\hsize]{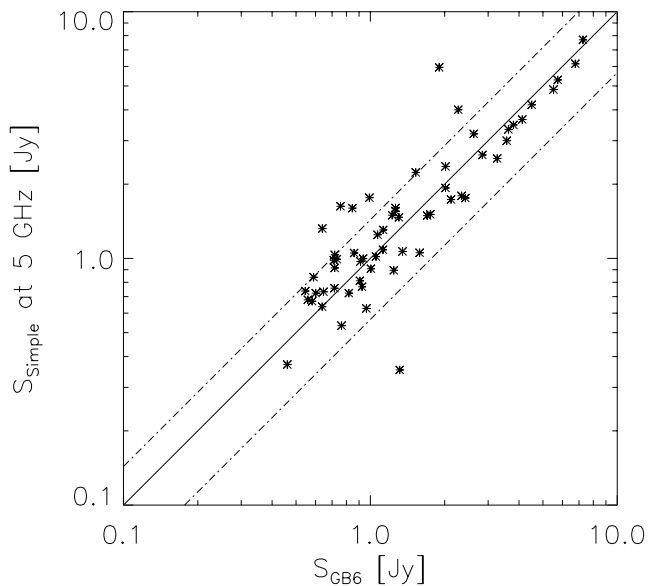}
 \caption{Comparison of SiMPlE and GB6 5-GHz flux densities. The solid line shows the best linear fit and the dot dashed lines enclose the area between the best fit plus and minus one standard deviation of the flux density ratio.
 }
\label{fig:cfr_GB6}
\end{figure}
For each source we have identified the best epoch of observations as the one with both 5 and 8.3 GHz observations. In case of multiple epochs with both frequencies we have chosen those having higher signal-to-noise ratio at both frequencies.
We thus have a sample of 61 sources for which we estimated the spectral indices $\alpha_{5}^{8.3}$ (defined according to the flux density being $\propto \nu^\alpha$). The median spectral index is -0.09 with a standard deviation of the distribution equal to 0.62 (see Fig. \ref{fig:alpha_5_8}). This is compatible with the 5-10~GHz spectral index found for the PACO bright sample (the selection was based on 20-GHz flux densities $>500$ mJy).

33 sources have counterparts within the {\it Planck} ERCSC (The {\it Planck} Collaboration 2011a). The median of the 5-30 GHz spectral index is -0.07 (see Fig. \ref{fig:alpha_30_70}).
As elsewhere stated (Massardi et al. 2008, 2009, 2011) the bright high-frequency selected samples are mostly constituted by flat spectrum sources in this frequency range (see also The {\it Planck} Collaboration 2011c, d).

For this sample in the ERCSC the 30-70 GHz spectral index is -0.16. As compared with the 100-GHz flux density limited sample in Sadler et al. (2008), the steepening is more pronounced for brighter samples.

We have compared our sample flux densities with the 4.85 GHz GB6 catalogue (Gregory et al. 2006). 60 objects in the Northern sample have a counterpart in the GB6 catalogue (the remaining objects include the 10 sources for which we have performed maps of the surrounding region and the objects in the region with $\delta >75^\circ$ that is not covered by the GB6 survey). The flux density comparison is shown in Fig. \ref{fig:cfr_GB6}. Performing a linear fit of these data we found the following relation between the flux densities from SiMPlE and GB6: 

\begin{equation}
S_{SiMPlE, 5}/{\rm [Jy]}=(0.900\pm0.057)\cdot S_{GB6}+(0.250\pm0.130).
\end{equation}

For this wide sub-sample the rms of the fractional displacement is 43 per cent: since more than 14 years divide the two catalogues, this effect is probably entirely due to source variability. This is comparable with the 38 per cent variability found, at the same frequency on few years timescales, on the PACO sample confirming earlier indications that the variability increases with the time lag, for lags of up to several years. Our results seem to indicate that the asymptotical value for the source variability at 5 GHz is $\sim 43 per cent$.

\section{Summary}
\label{sec:conclusions}

In this work, we have described the sample selection, the observing strategy, the data reduction procedures, and the software tools developed for the Simultaneous Medicina {\it Planck} Experiment (SiMPlE) project.
The selected sample includes 253 sources in the NEWPS sample and 11 candidate variable Galactic sources.

We have presented some tools that we have developed to analyse data collected with the Medicina 32-m single dish.  The tools allow us to efficiently schedule source lists, to visually inspect the scan quality, to compose and analyse maps of sky patches, and to calibrate and estimate the source flux densities.

The developed pipeline allowed us a fast and detailed automated analysis of our entire data-set, and the whole software package can be easily adapted for analogous OTF scan-based experiments.

As a result we have shown the results obtained for a sub-sample of 79 sources with $\delta \geq 45^{\circ}$ observed in 22 runs in the June-December 2010 epoch, at 5 and 8.3 GHz. Data for 2 of them have been flagged for bad weather conditions. For 10 targets we have performed maps of the surrounding sky area confirming the detection in the WMAP maps for 4 of them. These results confirm that the NEWPS catalogue could be considered 95 per cent reliable. 

The comparison with the ERCSC and the analysis of the spectral behaviour have confirmed that bright sample sources are mostly composed by flat-spectrum objects up to 30 GHz, but show an overall steepening in the spectrum in the frequency range  30-70 GHz. The comparison of the 60 sources in common with the GB6 catalogue show a rms fractional displacement due to source variability of 43 per cent that is an indication of the long-term variability for bright sources on time scales of more than 10 years.

In addition to the SiMPlE Northern Sample, we also report the observations for a list of 10 sources with $\delta < 0^{\circ}$ chosen from the targets of the PACO project. The comparison of the flux densities obtained with the two instruments and the SEDs of these targets showed that the ATCA and the Medicina radio telescopes have a consistent flux density scale, despite the fact that the calibration and the data reduction follow independent methods. Larger samples are being observed simultaneously with the two telescopes to confirm this finding.

Observations for the SiMPlE experiment are continuing for the first semester of 2011 in order to overlap with at least 2 full surveys of the {\it Planck} satellite and exploiting also the commissioning phase of the 22-GHz multi-feed receiver at the Medicina radio telescope. The SiMPlE sample might constitute a reference sample for this new telescope and for on-going Northern Hemisphere surveys.

\section*{Acknowledgments}
We acknowledge financial support for this research by ASI (ASI/INAF Agreement I/072/09/0 for the {\it Planck} LFI activity of Phase E2 and contract I/016/07/0 'COFIS').
Based on observations with the Medicina telescope operated by INAF - Istituto di Radioastronomia. We gratefully thank the staff at the Medicina radio telescope for the valuable support they provide. A particular acknowledgment goes to Andrea Orlati and Alessandro Orfei.
We warmly thank Uwe Bach and Alex Kraus of the 100-m Effelsberg telescope of the Max Planck - Institut f\"{u}r Radioastronomie who have provided important feedback on calibrator flux densities prior to publication. It is a pleasure to thank Gianfranco De Zotti, Luigina Feretti, and Nazzareno Mandolesi for constructive and stimulating conversations.



\bsp

\newpage

\setcounter{table}{3}
\begin{table*}
\begin{center}
\caption{A sub-set of the SiMPlE Northern sample. The table will be replaced by the complete sample.}
\label{tab:catalogue}
\begin{tabular}{lccccccccc}
\hline
ID & RA& DEC & Date &Sim & $S_{5GHz}$ & $\sigma_{5GHz}$& $S_{8.3GHz}$ & $\sigma_{8.3GHz}$ & ID\\
&[h]&[deg]&&&mJy&mJy&mJy&mJy&\\
\hline
      n513&23:56:56.609& 67:51:37.705&  2010-07-25&.&    -&   478&  -& 145.2& 87GB235426.3+673455\\
      n513&23:56:56.609& 67:51:37.705&  2010-09-04&*&   371&    -&  92.9&  -& 87GB235426.3+673455\\
      n512&23:54:21.724& 45:53:04.401&  2010-07-18&*&  1369&    -& 146.8&  -&       GB6J2354+4553\\
      n512&23:54:21.724& 45:53:04.401&  2010-07-24&*&  1304&    -& 102.0&  -&       GB6J2354+4553\\
      n512&23:54:21.724& 45:53:04.401&  2010-07-25&*&    -&  1925&  -& 226.3&       GB6J2354+4553\\
      n512&23:54:21.724& 45:53:04.401&  2010-08-08&*&    -&  1143&  -& 101.4&       GB6J2354+4553\\
      n511&23:56:22.822& 81:52:52.604&  2010-07-18&.&   877&    -& 152.1&  -&  NVSSJ235622+815252\\
      n511&23:56:22.822& 81:52:52.604&  2010-07-24&.&  1112&    -& 112.6&  -&  NVSSJ235622+815252\\
      n511&23:56:22.822& 81:52:52.604&  2010-08-08&.&    -&   757&  -& 121.2&  NVSSJ235622+815252\\
      n511&23:56:22.822& 81:52:52.604&  2010-10-08&*&    -&   666&  -& 107.5&  NVSSJ235622+815252\\
      n497&23:22:26.001& 50:57:51.996&  2010-07-24&*&  1467&    -& 106.3&  -&  NVSSJ232226+505752\\
      n497&23:22:26.001& 50:57:51.996&  2010-07-25&*&    -&  1216&  -& 224.3&  NVSSJ232226+505752\\
      n497&23:22:26.001& 50:57:51.996&  2010-10-08&.&    -&  1252&  -& 102.9&  NVSSJ232226+505752\\
      n461&21:53:28.704& 47:16:03.007&  2010-07-25&.&    -&  1054&  -& 300.2&       GB6J2153+4716\\
      n461&21:53:28.704& 47:16:03.007&  2010-10-10&.&  1321&    -&  73.8&  -&       GB6J2153+4716\\
      n437&20:38:37.009& 51:19:13.098&  2010-07-18&.&    -&  2664&  -& 240.9&       GB6J2038+5119\\
      n437&20:38:37.009& 51:19:13.098&  2010-07-25&.&    -&  2805&  -& 247.4&       GB6J2038+5119\\
      n437&20:38:37.009& 51:19:13.098&  2010-08-02&.&    -&  2640&  -& 251.3&       GB6J2038+5119\\
      n437&20:38:37.009& 51:19:13.098&  2010-08-08&.&    -&  2703&  -&  86.3&       GB6J2038+5119\\
      n437&20:38:37.009& 51:19:13.098&  2010-09-30&.&  2423&    -& 218.2&  -&       GB6J2038+5119\\
      n433&20:22:06.702& 61:36:58.895&  2010-07-18&*&  3197&  3057& 181.9& 216.1&       GB6J2022+6137\\
      n433&20:22:06.702& 61:36:58.895&  2010-07-25&*&    -&  3492&  -& 275.6&       GB6J2022+6137\\
      n433&20:22:06.702& 61:36:58.895&  2010-08-02&*&    -&  3113&  -& 259.8&       GB6J2022+6137\\
      n433&20:22:06.702& 61:36:58.895&  2010-08-08&*&    -&  3164&  -&  90.5&       GB6J2022+6137\\
      n433&20:22:06.702& 61:36:58.895&  2010-08-25&*&    -&  3030&  -& 321.3&       GB6J2022+6137\\
      n433&20:22:06.702& 61:36:58.895&  2010-09-30&.&  3421&    -& 228.7&  -&       GB6J2022+6137\\
      n430&20:09:52.507& 72:29:19.488&  2010-07-18&.&   889&    -& 186.8&  -&       GB6J2009+7229\\
      n430&20:09:52.507& 72:29:19.488&  2010-07-24&.&   811&    -& 114.8&  -&       GB6J2009+7229\\
      n430&20:09:52.507& 72:29:19.488&  2010-07-25&.&    -&   953&  -& 253.4&       GB6J2009+7229\\
      n430&20:09:52.507& 72:29:19.488&  2010-09-30&*&  1205&    -& 243.8&  -&       GB6J2009+7229\\
      n429&20:05:31.289& 77:52:43.898&  2010-07-18&.&  1458&  1193& 180.6& 240.9&  NVSSJ200531+775243\\
      n429&20:05:31.289& 77:52:43.898&  2010-07-25&.&    -&   951&  -& 172.1&  NVSSJ200531+775243\\
      n429&20:05:31.289& 77:52:43.898&  2010-08-02&.&    -&  1016&  -& 223.5&  NVSSJ200531+775243\\
      n429&20:05:31.289& 77:52:43.898&  2010-08-08&.&    -&  1477&  -& 123.5&  NVSSJ200531+775243\\

\hline
\end{tabular}
\end{center}
\end{table*}
\setcounter{table}{3}

\label{lastpage}

\end{document}